\documentclass[epj]{svjour}

\usepackage{graphicx}
\usepackage{fancyhdr}
\usepackage{epsfig}
\usepackage{colordvi}
\usepackage{axodraw}

\def\mytitle{My title} 
\def\myauthors{My name}  
\def\mytype{My type of session}
\def\mysession{My session}
\def\mytitle{From Symmetry to Supersymmetry} %Put your title here!
\def\myauthors{Julius Wess}    %Put your name here!
\def\mytype{Review}
\def\mysession{\myauthors}
\begin{document}
\title{From Symmetry to Supersymmetry}
%\subtitle{\emph{*This text is an unfinished draft of J.Wess' contribution to these Proceedings}}
\author{Julius Wess\inst{1}\inst{2}
\thanks{\emph{This text is the draft of Julius Wess' contribution to the Proceedings of SUSY07 (KIT Karlsruhe)
              and to ``Supersymmetry at the dawn of the LHC'' in Eur. Phys. J. C59/2. The manuscript, which Wess could 
              not finish before his death, has been edited for the publication by I.~Gebauer and P.M.~Zerwas.}}%
}                     % Do not remove
\institute{Sektion Physik der Ludwig-Maximilians-Universit\"{a}t, Theresienstr. 37, D-80333 M\"{u}nchen
\and Max-Planck-Institut f\"{u}r Physik (Werner-Heisenberg-Institut), F\"{o}hringer Ring 6, D-80805 M\"{u}nchen
}
\date{}
\abstract{% 
\it
 "...die Wirklichkeit zeigt sich in unseren Erlebnissen und Forschungen nie anders wie durch ein Glas, das teils den Blick durchl\"{a}sst, teils den Hineinblickenden widerspiegelt."
% \vspace
\\
\\
 R. Musil "Der Mann ohne Eigenschaften", Zweiter Teil, Kapitel "Das Sternbild der Geschwister oder Die Ungetrennten und Nichtvereinten"
\PACS{
      {PACS-key}{discribing text of that key}   \and
      {PACS-key}{discribing text of that key}
     } % end of PACS codes
} %end of abstract
\maketitle
%DO NOT REMOVE THIS LINE
%

It is quite easy to speak about symmetries, on one side. Everybody has a notion of symmetry, it is a very deeply rooted and widespread concept, ranging from art to science. In some way or another symmetry is perceived by everybody. I think it is worth mentioning that about thirty years ago there was strong interest in experimenting with apes to see how much they were able to learn. One objective was to see how apes would learn to paint. In one of these experiments one dot was made at one side of a piece of paper and the ape would then try to make a dot on the other side to balance it symmetrically. That's exactly what we are doing in physics.

It is quite difficult to speak about symmetries, on the other side. Everybody has his own concept of it, also physicists, and you never know if we communicate about the same thing.

Fortunately, mathematics with its strong capability to abstract has abstracted the concept of symmetries to the concept of groups. Group theory and the theory of representations of groups incorporate many of the different aspects of symmetries as we meet them in Nature, in art, in science etc. When referring to symmetry I mean it in the framework of group theory, representation theory, algebra and differential geometry.

In physics symmetries have been used all along. Group theoretical methods make it much simpler to get information about a system. A problem like the Kepler problem, for example, would be much harder to solve without using rotational symmetry.

Through symmetries you might get information \\about a system without really understanding the physical laws that govern it or without being able to solve the dynamical laws if you know them. You may think about the trivial example of a scale. Without knowing the theory of gravitation you are quite convinced that based on the symmetry of a scale the weight on the left hand side and on the right hand side of the scale is the same. This demonstrates how symmetries can help us to find our way through a system without really knowing its laws.

In addition symmetries have a very strong interplay with experimental physics via conservation laws. Conservation of energy, momentum and angular momentum can be measured experimentally. They are linked to the invariance under time translation, space translation and space rotation. Noether's theorem states it very precisely: if we know that a system is invariant under some symmetry transformations then we can show that there are corresponding conservation laws and we know how to find their explicit form. We know that there is a conservation law of the electric charge - we ask for a symmetry and find it in the phase transformation of the Schr\"{o}dinger wave function. This shows that there is a very strong connection between abstract mathematical concepts and experimental facts. We are very lucky to have such an interplay in physics.

Symmetries in modern physics have taken an even stronger role to such an extent that the laws of modern physics cannot even be formulated without the concept of symmetries. To make the framework of local quantum field theory meaningful, symmetries have to be invoked from the very beginning. It is not that we know the laws and try to find their symmetries, but rather we have to implement the symmetries from the very beginning to be able to formulate these laws in a meaningful way.

\begin{figure}
\begin{center}
\begin{picture}(360,350)(0,20)
\SetColor{Black}
\DashLine(230,160)(10,160){10}
\Line(110,340)(110,20)
\Text(110,360)[ct]{Symmetries in particle physics}
\Text(50,330)[ct]{Space-time}
\Text(170,330)[ct]{Inner space}
\Text(50,300)[ct]{Translations}
\Text(170,300)[ct]{Phase transformations}
\Text(50,270)[ct]{Rotations $SU(2)$}
\Text(170,270)[ct]{Isospin $SU(2)$}
%\Text(50,240)[ct]{Rotations $SU(2)$}
\Text(170,240)[ct]{$SU(3)$}
\Text(50,210)[ct]{Lorentz group}
\ArrowLine(50,200)(50,110)
\Text(170,210)[ct]{$SU(3)_c \times SU(2) \times SU(1)$}
\ArrowLine(170,200)(170,110)
\Text(50,150)[ct]{Conformal group}
\Text(170,150)[ct]{GUT $SU(5)$}
\Text(50,100)[ct]{Gravity}
\Text(170,100)[ct]{Standard Model}
\Text(50,70)[ct]{long distances}
\Text(170,70)[ct]{short distances $\sim 10^{-16}cm$}
\Text(50,40)[ct]{Classical Theory}
\Text(170,40)[ct]{Quantum Field Theory}
\end{picture}
\caption{Symmetries in particle physics}
\label{fig:1}       % Give a unique label
\end{center}
\end{figure}
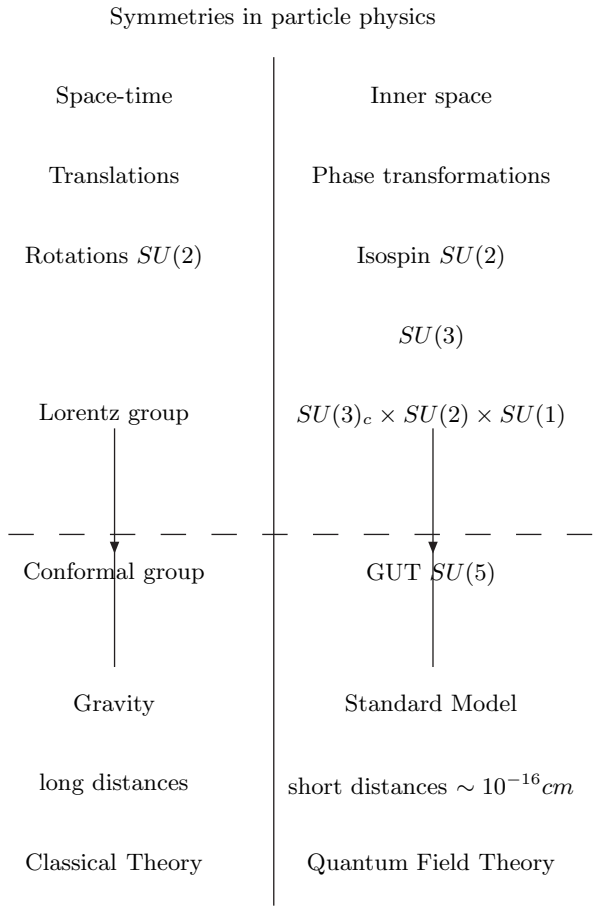

I have prepared a figure listing some of the fundamental symmetries (Figure \ref{fig:1}). It shows two separate columns: one, on the left, for space-time symmetries and one, on the right, for symmetries in an inner space.

At the top of the left column we have the rotation group. It is the group from which a physicist can learn what group theory is about. We know that the laws of electrostatics and magnetostatics are invariant under the rotation group and that Newton's laws allow rotational invariance as well. The rotation group is strongly related to SU(2), the special unitary group in two dimensions, which is connected to the concept of spin. You know all this from quantum mechanics. When electrostatics and magnetostatics are combined to Maxwell's theory of electromagnetism we encounter the Lorentz group. Though electrostatics and magnetostatics describe forces that are different in strength by the order of magnitude of the velocity of light, they nevertheless are part of the same theory. We could say that the unification comes about by generalizing the rotation group to the Lorentz group. This is a good model and yields a good understanding of how theories can be unified by enlarging a group.

The right column shows symmetries in an inner space. I have mentioned the phase transformation of the wave function. Wave functions can span a higher-dimensional space and this is the space which we call inner space. The first step was done by Heisenberg. He had learned what $SU(2)$ is in the framework of the rotation group and spin and put this concept to work in the inner space as isospin. This was later generalized to $SU(3)$. A successful attempt, which we now understand in terms of the quark model. A very successful model, the Standard Model of particle physics, is based on a group $SU(3)_{C}$ for colour, $SU(2)_{W}$ and $U(1)$ for weak and electromagnetic interactions . The Lorentz group together with this \\$SU(3)_{C} \times SU(2)_W  \times U(1)$ is at the basis of the Standard Model. This model describes physics very well as we know it in our laboratories, down to a scale of $10^{-16}cm$.

It is natural to ask if the $SU(3)_C \times SU(2)_W \times U(1)$ symmetry is not part of a larger group, $SU(5)$, for instance. This then gives rise to a theory where strong, weak and electromagnetic interactions would be truly unified. Such a theory we call GUT - for Grand Unified Theory - but what Nature knows about this, we do not yet know. It is the Standard Model based on the Lorentz group and the group $SU(3)_C \times SU(2)_W \times U(1)$ that is supported and by now well tested by experiment.

Looking at the two columns it seems that Nature (or we) has used the same concept of symmetry twice. But has Nature chosen it separately? Has it done the same thing twice? It is natural to ask for a bridge between the two columns.

We have learned from the Maxwell equations that one can make a transformation parameter space-time dependent. This is the property of the gauge transformation in Maxwell theory. We can identify this parameter in Maxwell theory with the parameter of the phase transformation of the Schr\"{o}dinger wave function and build a gauge theory, as we call it today. Thus the idea of a gauge theory is born. You demand that the theory should be invariant under a group that acts in inner space and has space-time dependent parameters. This is the concept on which the Standard Model is built. The Standard Model is the gauge theory for the group $SU(3)_C \times SU(2)_W \times U(1)$. If the same idea of gauging is used for the space-time symmetry, the Lorentz group, one arrives at Einstein's theory of gravitation.

We know that the Standard Model can be interpreted -- via the concept of renormalization -- as a Quantum Field Theory and as such it is experimentally extremely successful at short distances, the scale of $10^{-16}cm$ I have mentioned before.

We would have liked to find deviations from the Standard Model experimentally to get a hint where to go next: $SU(3)_C \times SU(2)_W \times U (1)$ seems to be the simplest choice one can make and it works and it works and all the experiments verify it again and again. Is Nature not more sophisticated? Doesn't it know about GUT?

As for Einstein's gravitational theory we know that it is a very good theory for long distances. We have no reason to doubt its validity there. Our understanding of space-time at large distances is based on this theory.

We have the situation that the Standard Model and Einstein's theory of gravitation describe the data observed in the laboratory as well as in astronomy and astrophysics very well. Even cosmology based on these two theories is very reasonable.

Looking at our columns we now have an even more puzzling situation. The same concept of symmetries and gauging them gives in the left column a very good classical theory of gravity, defending itself against quantization by an abundant number of singularities. There seems to be a deep conflict between the classical theory of gravitation and quantum field theory. On the other hand the same idea about symmetry and gauging in the right column leads to a very good mathematically and experimentally successful model of a renormalizable quantum field theory.

Is gauging all or is there a deeper connection between space-time and inner space symmetries? In nuclear physics, a long time ago, Wigner and Hund %\cite{wigner} 
proposed a group SU(4) that has SU(2) of spin and SU(2) of isospin as a subgroup. This way they unified space-time and inner space and got a good classifiation of nuclear levels. At the time of SU(3) in particle physics this idea was generalized to SU(6), incorporating SU(2) of spin and SU(3) of inner space. Reasonable experimental predictions about masses of particles with different spin - spin 0 and spin 1 as well as spin 1/2 and spin 3/2 - can be based on SU(6). This is now better understood on the basis of the quark model. As the quark model was not yet known at the time, many attempts were made to extend the SU(6) model to incorporate the Lorentz group. But this proved to be impossible. It was impossible to build a Lorentz invariant model that at low energies would have SU(4) or SU(6) as a symmetry. If a lot of physicists try something and it does not work then it might be clever to try to prove that it cannot work. And this is what was done. It started with work by O'Raifeartaigh \cite{Raif} and came to a very elegant formulation which now is known as the Coleman Mandula "no-go" theorem \cite{Cole}.

This theorem tells us that for a theory in four dimensions with the Lorentz group as a symmetry group and satisfying a certain number of axioms, which I am going to tell you about in a minute, the only possibility for a symmetry group is the direct product of the Lorentz group with some compact inner group.

Surprisingly enough we have a theorem that separates the two columns on the basis of very fundamental axioms.

Now some remarks on the axioms. Apart from the Lorentz invariance the axioms state that it should be a theory based on quantum mechanics and that it should be local, it should be a local quantum field theory. Local here means local in the microscopic sense. Two measurements that are separated space-like cannot influence each other, no matter how small the distance. The locality of the theory is based on the locality of the fields. In addition we assume that there is a unique lowest energy state, a ground state which we call vacuum, and that all the other states have larger energy. Probability is supposed to be conserved in the quantum mechanical sense - and finally we assume that there is only a finite number of different particles. These seem altogether very reasonable assumptions, but they have as a consequence that you cannot combine the two columns as we tried by postulating an SU(4) or an SU(6) symmetry.

There is one problem with this set of axioms. With the exception of the free field theory we do not know a local quantum field theory the existence of which we can rigourously prove and which satisfies all the axioms. We have invented very powerful methods of perturbation theory, of separating infinities, going through a renormalization scheme. We are able to extract in this way information that we can test experimentally. In some way it is a kind of art, but it works beautifully. On this basis we understand the models I have been talking about and we make them successful. This is, however, a formulation of a model a mathematician would not like to accept as a theory. But the success in comparing it with experimental data is so strong that we cannot dismiss this type of theories.

Now I would like to make a point. If we try to build a theory on the setting just discussed and we want to have spin zero, spin one and spin one-half fields, we start from a free field theory and try to arrange the multiplets and couplings such that there should be only a finite number of divergences in perturbation theory. We start with the tree level, study the energy momentum dependence of the Feynman diagrams and arrange the model in such a way that this dependence is as smooth as possible in order to facilitate integration over energy and momentum variables at the one loop level. The result is a gauge theory with spontaneous symmetry breaking. Thus, without knowing about group theory and symmetries the whole concept of gauge theories could be invented by a good physicist on dynamical grounds. This was shown about twenty years ago by Cornwell, Llewellyn Smith and Steve Weinberg %\cite{Corn}
. They pioneered this approach. After you know this result it is much easier to take a textbook on group theory, formulate your model based on your knowledge of group theory, and gauge the symmetry group. In this way you obtain a model and you will find it to be a renormalizable quantum field theory. This can be proved from gauge invariance. Noether's theorem and the conserved currents are at the heart of this proof. The current has to be a well-defined object and it helps to relate infinite and, therefore, undefined constants such that there is only a finite number left at the end. This defines the renormalization scheme. This is what I meant in the beginning when I said that even in formulating a physical theory we have to use the concept of symmetry. But this also raises the question whether symmetries are very basic or whether they can be derived from the dynamics of the system - based on some reasonable axioms.

Our way of thinking is very much influenced by the deep and widespread notion of symmetry to an extent that if we find a beautiful symmetry in a dynamical system we say that we understand this system. If there is a deviation from the symmetry we say: there must be something going on which we do not understand.

There is another good reason to start from symmetries instead of dynamical requirements. The second way would be difficult and, let me say, ugly. Mathematicians have not developed concepts along this ill-defined line. Thus, we have no mathematical machinery that we can use. Contrary to the first approach, where we have all this beautiful mathematics.

There is a surprising way of combining space-time symmetries with internal symmetries, and this is by supersymmetry. This is achieved by generalizing the concept of symmetry. As you know, symmetries can be formulated in terms of commutator relations, as the commutator relations of angular momentum in quantum mechanics. A large class of groups - the Lie groups - are related to Lie algebras that are defined by commutator relations. All the symmetries we have mentioned are of this type.

From Dirac we know that not only commutators but also anticommutators are a very useful concept, especially when we deal with particles with half integer spin.

The idea is to generalize the concept of symmetries to a structure which is formulated in terms of commutators and anticommutators as well. This is not a new idea in mathematics, such structures - graded Lie algebras - have been thoroughly investigated e.g. by Berezin %\cite{Ber}
. But can such a concept be realized in a quantum field theory? The answer is yes. There is a quite unique symmetry and its uniqueness is based on the no-go theorem of Mandula and Coleman (Figure \ref{fig:2}).

\begin{figure}
\begin{center}
\begin{picture}(360,200)(0,20)
\SetColor{Black}
\Line(70,120)(120,180)
\Text(90,150)[rt]{Space-time}
\Line(170,120)(120,180)
\Text(150,150)[lt]{Inner space}
\Text(120,130)[ct]{no go theorem}
\Text(120,110)[ct]{based on local field theory in four dimensions}
\Text(120,100)[ct]{\bf $\Downarrow$}
\Text(120,90)[ct]{supersymmetry}

\end{picture}
\vspace*{-2cm}
\caption{}
\label{fig:2}       % Give a unique label
\end{center}
\end{figure}
This was shown by Haag, Lopuszanski and Sohnius \cite{Haa}.
Here the theorem does not tell you that it does not go, it tells you that it goes in a very unique way expressed by the formula:
\begin{equation}
\lbrace Q^{N}_{\alpha},{\bar{Q}}^{M}_{\alpha}\rbrace_{+}=2{\gamma}^{\nu}_{\alpha \beta} P_{\nu} \delta^{N,M}.
\end{equation}

The charges $Q^{N}_{\alpha}$ are spinorial charges, $\alpha$ is a spinor index and $Q^{N}_{\alpha}$ is a Majorana spinor. $N$ is a free index. If $N$ goes from one to two we speak of $N = 2$ supersymmetry, if it goes from one to four we speak of $N = 4$ supersymmetry. The four-vector $P_{{\nu}}$ is the energy momentum four-vector that generates the translations in space-time. The structure constants of this algebra are the Dirac $\gamma$ matrices and the Kronecker symbol. This algebra can be combined with the algebra of the Lorentz group. This, then, is the algebra of supersymmetry.

 These charges can now be realized in terms of local fields and the algebraic relations of supersymmetry hold on the basis of the canonical commutation relations of the fields. The charges are related to currents, which are 3/2 currents in the same way as energy momentum is related to the energy momentum density tensor which is a spin 2 object. If the theory is gauged, the currents are the sources of fields - for the electric current this is the photon, with spin 1, for the energy momentum tensor it is the graviton, with spin 2, and for the supercurrent it is the gravitino with spin 3/2.

 Supersymmetric theories have a very encouraging property. As quantum field theories they are less divergent as they would be without supersymmetry.

 Supersymmetric theories have also a very discouraging property. It follows directly from the algebra that a supersymmetric theory has to have an equal number of bosonic and fermionic degrees of freedom degenerate in mass. This is not how Nature is. We have fermions and bosons but not in the same multiplet structure and in no way degenerate in mass.

 The two properties lead to frustration, the more so as they are not independent. Let me explain this with a simple example. Take a bosonic and a fermionic harmonic oscillator, both with the same frequency. The zero point energy will have opposite sign and adding them leads to the cancellation of the zero point energy. A field can be viewed as an infinite sum of harmonic oscillators. The zero point energy will add up to an infinite vacuum energy except in a theory where the bosonic and fermionic contributions cancel. This was already known to Pauli but he also knew that the world is not like this. Supersymmetry relates that cancellation to an algebraic structure of the theory and you might be led to believe that it is now based on a deeper property of the theory and Nature. Since the days of Pauli we have learned to deal with symmetries that are spontaneously broken. The field theoretic properties of such theories with spontaneously broken symmetries are maintained, but at a phenomenological level at low energies the symmetry appears to be broken by sizable effects.

 The same mechanism that leads to the cancellation of the vacuum energy leads to many other cancellations of divergences. These improved renormalization properties of the theory can be traced back to the cancellation properties of diagrams with bosonic or fermionic internal lines. It can be shown that there are parameters in the theory which do not get any radiative correction, not even a finite one. If not a miracle, this is a sensation in a quantum field theory. On the non-quantized level you can introduce parameters like certain masses or couplings that are not changed by radiative corrections at all. Naturally, this has consequences for the particle phenomenology based on supersymmetric theories.

 Let me first discuss the fact that supersymmetric theories have an equal number of bosons and fermions.

 In Nature we know quarks and leptons to be fermions. These are the particles that constitute matter. Each of these fermions has to have two bosonic partners as each spin one half particle has two degrees of freedom. We do not know such partners in Nature but we can give them names. The SUSY partners of the quarks we call squarks and the SUSY partners of the leptons we call sleptons. The "s" stands for scalar of the supersymmetric partner.

 In Nature we know the photon, the vector particles of weak interaction, the gluons, the graviton and the Higgs to be bosons. These are the particles that constitute the forces. Each of these bosons has to have a fermionic partner. We give them names: photino, wino, zino, gluino, gravitino and Higgsino.

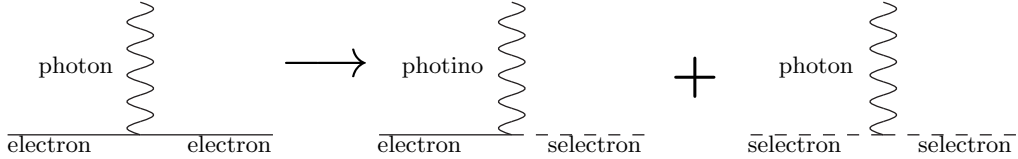
\begin{figure*}
\begin{center}
\begin{picture}(360,200)(0,20)
\SetColor{Black}
\Photon(50,120)(50,170){5}{5}
\Text(40,150)[rt]{photon}
\Line(0,120)(50,120)
\Text(0,120)[lt]{electron}
\Line(50,120)(100,120)
\Text(100,120)[rt]{electron}
\Text(120,150)[ct]{\huge \bf $\longrightarrow$}

\Photon(190,120)(190,170){5}{5}
\Text(180,150)[rt]{photino}
\Line(140,120)(190,120)
\Text(140,120)[lt]{electron}
\DashLine(190,120)(240,120){5}
\Text(240,120)[rt]{selectron}
\Text(260,150)[ct]{\huge \bf +}

\Photon(330,120)(330,170){5}{5}
\Text(320,150)[rt]{photon}
\DashLine(280,120)(330,120){5}
\Text(280,120)[lt]{selectron}
\DashLine(330,120)(380,120){5}
\Text(380,120)[rt]{selectron}
\end{picture}
\vspace*{-3cm}
\caption{Supersymmetric vertices in QCD.}
\label{fig:3}       % Give a unique label
\end{center}
\end{figure*}

 With these particles we can build models and consider it a success that we know already half of the particles in such models. But we know also their couplings, which are entirely determined by the couplings of the particles we know. Take a Feynman diagram with the known particles (Figure \ref{fig:3}). A line in such a Feynman diagram that either goes from an incoming particle to an outgoing particle or that forms a closed loop can and has to appear in an equivalent diagram where it is replaced by a line that is associated with SUSY partners. Doing this for all the lines you obtain all the diagrams of a certain supersymmetric theory. The coupling constants on the respective vertices are the same as in the theory you started from.

 In a truly supersymmetric theory the masses would also be the same. Not so if the symmetry is spontaneously broken. We know that for spontaneously broken gauge symmetries the mass difference within a multiplet might be quite big -- like the mass difference between the photon and the W or Z.

 It is possible to break supersymmetry as well in such a way that its renormalization properties remain valid, the mass difference between SUSY partners can then be quite big -- we call it the SUSY gap.

 The radiative correction, however, will not completely cancel - we can expect finite contributions which then will also be in the order of the SUSY gap. Among the masses and coupling constants of a gauge theory that will have this property are the Higgs masses and the Higgs couplings.

Knowing this one can relate this property to another phenomenon in particle theory. Assume that there is a GUT theory. The unifying gauge group has to be spontaneously broken to render the Standard Model as we know it. The symmetry group \\$SU(3)_C \times SU(2)_W \times U(1)$ of the standard model is then spontaneously broken to the theory of electromagnetism and weak interactions as we know it at low energies.

 This breaking of the symmetries is triggered in a quantum field theoretical model by parameters, the Higgs masses and Higgs couplings, which have to be renormalized in a non-supersymmetric theory by an infinite amount. It is very difficult to understand that the breaking scheme in the two sectors of the breaking is stable and respects the scale. This is the hierarchy problem in particle physics. If, however, the theory is a spontaneously broken supersymmetric theory, then the relevant parameters obtain only radiative corrections of the order of the SUSY gap. If the SUSY gap is of the order of the electroweak breaking scale characterized by the W mass then we would understand the stability of the breaking of the symmetries in a GUT theory. In such a scenario the SUSY gap has to be of the order of one TeV - an order which will be accessible to experiments soon.

My personal belief is that it would be a waste having such a beautiful symmetry as supersymmetry just to solve the hierarchy problem. Supersymmetry might play a much more fundamental role at higher energies and solving the hierarchy problem would just be one of the lower energy remains of supersymmetry.

But let me stress that there are two inputs in this prediction - SUSY and GUT. In any combination they might be right or wrong.

 Now back to the theme mentioned already in the context of gauge theories. We see that supersymmetry has a very strong influence on the dynamical behaviour of a quantum field theory when it is treated in the framework of renormalization theory. We can start from this framework and ask for a model, e.g. with spin 0 and spin 1/2 particles, that behaves as smoothly as possible not only on the tree level but on the one-loop level as well. We could ask for the cancellation of the infinities in the vacuum energy, in this way we would establish an equal number of fermionic and bosonic degrees of freedom. If we then asked for the non-renor\-ma\-li\-za\-tion properties about which I have been talking before we would construct a supersymmetric theory -- with, what we call possible soft breaking, a symmetry breaking that does not affect the renormalization behaviour. If we asked for the absence of any radiative correction we would obtain a strictly supersymmetric theory.

 So again it would be a dynamical concept, now put forward to the level of one-loop diagrams which would have led to the invention of supersymmetry, without knowing about any algebra, about groups or graded groups, just being a good physicist knowing how to handle Feynman diagrams. Somehow supersymmetry is the next logical step after gauge theory in the framework of renormalizable quantum field theories. You go from the tree level corresponding to a classical theory to the quantized level represented on the one-loop level, apply the same idea once more, and you arrive at supersymmetry. What we do not know is if Nature knows about this way of thinking or if Nature has a different logic, does things differently.

Now to the history of supersymmetry. It started with the work of Golfand and Likhtman \cite{3A}
. They thought about adding spinorial generators to the Poin- car\'{e} algebra, in that way enlarging the algebra. This was about 1970 and they were really on the track of supersymmetry. I will come back to this idea at the end of this talk as I think that this is the right question: can we enlarge the algebra, the concept of symmetry, by new algebraic concepts in order to get new types of symmetries?
Then in 1972 there was a paper by Volkov and Akulov \cite{Volk} which argued on the following line. We know that with spontaneously broken symmetries there are Goldstone particles, supposed to be massless. In Nature we know spin one half particles that have, if any, a very small mass, these are the neutrinos. Could these fermions be Goldstone particles of a broken symmetry? Volkov and Akulov constructed a Lagrangian, a non-linear one, that turned out to be supersymmetric. Of course today we know from Haag, Lopuszansky and Sohnius that it had to be supersymmetric. But being nonlinear, just as the nonlinear sigma model, the Lagrangian is highly non-re\-nor\-ma\-li\-za\-ble and does not show any sign of the renormalization properties which we now find so useful and intriguing in supersymmetry.

Another path to supersymmetry came from two-dimensional dual models. 
Neveu and Schwarz et al. \cite{3B} 
had constructed models which had spinorial currents related to supergauge transformations that transform scalar fields into spinor fields. The algebra of the transformation, however, only closed on mass shell. The spinorial currents were called supercurrents and that is where the name "supersymmetry" comes from.

In 1973 Bruno Zumino and I published a paper \cite{3C} 
where we established supersymmetry in four dimensions, constructed renormalizable Lagrangians and exhibited non-re\-nor\-ma\-li\-za\-tion properties on the one-loop level. Our starting points were the supercurrent and the strong belief in Noether's theorem.

 Another paper by another author that could have led to supersymmetry based on the non-re\-nor\-ma\-li\-za\-tion properties of perturbative quantum field theories was never written.

 With supersymmetry it is very natural to extend the concept of space-time to the concept of superspace. Energy momentum generates translations in four-\\dimensional space-time, so it is natural to have the anticommuting charges generate some translations in an anticommuting space. This new space together with the four-dimensional Minkowsky space is called superspace.

\begin{figure}
\begin{center}
\begin{picture}(360,310)(0,20)
\SetColor{Black}

\Line(50,290)(80,260)
\Text(40,300)[lt]{$\Phi$}
\Line(110,290)(80,260)
\Text(120,300)[rt]{$\Phi$}
\Line(50,230)(80,260)
\Text(40,230)[lb]{$\Phi$}
\Line(110,230)(80,260)
\Text(120,230)[rb]{$\Phi$}
\CCirc(80,260){10}{Black}{White}
\Text(80,210)[bc]{\bf no radiative corrections}

\CCirc(80,170){10}{Black}{White}
\Text(140,170)[lc]{\bf : Vacuum-Vacuum}

\Line(40,120)(80,120)
\Text(30,120)[lc]{$\Phi$}
\Line(80,120)(120,120)
\Text(130,120)[rc]{$\Phi$}
\CCirc(80,120){10}{Black}{White}
\Text(140,120)[lc]{\bf $m$: mass-term}

\Line(80,70)(80,30)
\Text(80,80)[tc]{$\Phi$}
\Line(40,30)(80,30)
\Text(30,30)[lc]{$\Phi$}
\Line(80,30)(120,30)
\Text(130,30)[rc]{$\Phi$}
\CCirc(80,30){10}{Black}{White}
\Text(140,30)[lc]{\bf $\lambda$: coupling}
\end{picture}

\caption{Non-renormalization.}
\label{fig:4}       % Give a unique label
\end{center}
\end{figure}
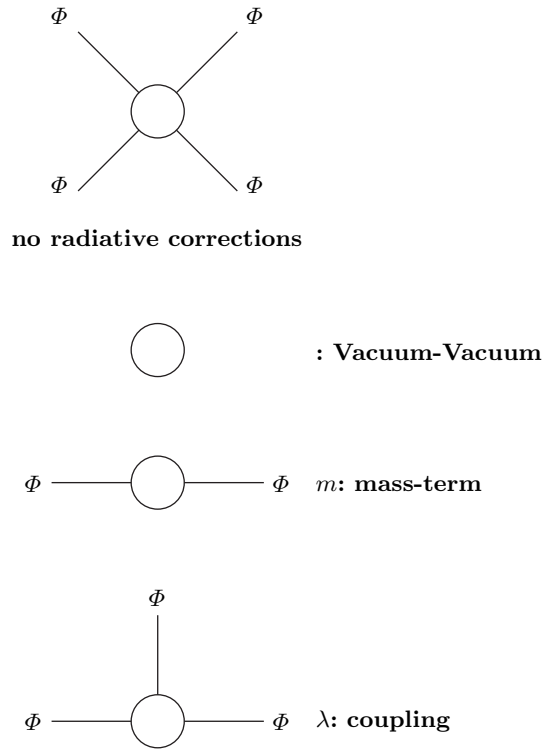

 Fields will now be functions of the superspace variables and they will incorporate SUSY multiples in a very natural way. This idea was pioneered by Salam and Strathdee \cite{Salam}
. Lagrangians can be formulated very elegantly in terms of superfields and the non-re\-nor\-ma\-li\-za\-tion theorems find very elegant formulations as well, as first shown by Fujikawa and Lang \cite{Fuji}.
The spin 0, spin 1/2 multiplets find themselves in the so-called chiral superfields ($\Phi$) and spin 1, spin 1/2 in the so-called vector superfields ($V$) (see Table \ref{tab:1}).
Any diagram in terms of superfields that has only external chiral fields but no conjugate chiral field $\Phi^*$ or vector field will not be renormalized (Figure \ref{fig:4}).

As no external field is in the class of chiral fields only we find the non-re\-no\-rma\-li\-za\-tion theorem for the vacuum energy. A short look at the superfield Lagrangian:
\begin{equation}
L=\Phi^{\dagger}e^{gV}\Phi+m^{2}\Phi^{2}+\lambda\Phi^{3}
\end{equation}
\begin{table}
\caption{}
\label{tab:1}       % Give a unique label
\begin{tabular}{lllllll}
\hline\noalign{\smallskip}
spin & 0 & $\frac{1}{2}$ & 1 & $\frac{3}{2}$ & 2 & \\
\noalign{\smallskip}\hline\noalign{\smallskip}
 & 2 & 1 & & & &$\Phi$, scalar multiplet \\
 &   & 1 & 1& & & $V$, vector multiplet \\ 
&   &  & & 1 & 1 & gravitational multiplet \\
\noalign{\smallskip}\hline
\end{tabular}
% Or use
\vspace*{1cm}  % with the correct table height
\end{table}
tells us that the mass terms of the chiral fields are of the form $\Phi^{2}$ and the coupling terms of the Higgs type are of the form $\Phi^{3}$. These couplings do not get renormalized. The kinetic term of a chiral field is of the form $\Phi^{*}\Phi$, there is a wave function renormalization.

 The superspace variables also play an important role if we want to gauge an internal symmetry of a supersymmetric theory. Gauging an internal symmetry means formulating a theory that is invariant under transformations with space-time dependent parameters. Space-time by itself is not a supersymmetric concept, we have to replace it by superspace. Gauging an internal symmetry in a supersymmetric theory means to formulate a theory that is invariant under transformations with superspace dependent parameters. This way we know how to supersymmetrize all the known gauge theories.

 Gauging supersymmetry as well finds a natural formulation in superspace. Einstein's theory of gravity can be formulated as a geometrical theory in four-dimensional space-time. Supergravity, the theory that has supersymmetry gauged, finds its formulation as a geometrical theory in superspace. Supergravity incorporates Einstein's theory because supersymmetry incorporates the Lorentz group. It improves the renormalization properties of the usual gravity theory, however, it does not make it a fully renormalizable theory. But it is closer to a dream to have also gravity as part of a renormalizable quantum field theory.

 All our interplay with symmetries has centered \\around the renormalization problem of perturbative quantum field theory. The singularities that have to be renormalized are consequences of the unsatisfying short distance behaviour of quantum field theories. Symmetries to some extent improve the situation. But is this really the way how Nature solves the problem of divergences at short distances?

 Another possibility is to loosen some of the axioms. But this has to be done in a very controlled way, first so as not to get into conflict with experimental facts, and secondly in order not to have the rules of the theoretical game too wide open and to turn theory into a book-keeping device.

 The only way up to now that meets these requirements is string theory. The concept of a point has changed to the concept of a string -- the axiom of locality has been loosened and there are infinitely many particles, the excitations of the string in the theory.

 At low energies the string picture might be compatible with our present experimental knowledge summarized in the Standard Model and Einstein theory of gravity. The theoretical framework of strings is again based on symmetries and differential geometry. Analyticity plays an important role as well. In string theory our concept of space-time is based on differential manifolds -- the closest relatives to flat space, however curved as they might be.

% In the spirit of Golfand we..
We could ask for a change in the algebraic structure of quantum field theory. We have based it on the canonical structure of quantum mechanics and adapted it to the algebraic possibilities of differentiable manifolds. Non-commutative differential geometry might give a mathematical frame that goes beyond differential manifolds. Attempts in this direction show that this could lead to a lattization of space-time at short distances. This opens possibilities worthwhile exploring. %Maybe Golfand was at the track of another theoretically successful development again.

\end{document}